\documentclass[aps,pra,twocolumn,amsmath,amssymb,showpacs,superscriptaddress]{revtex4-1}
\usepackage{graphicx}
\usepackage{dcolumn}
\usepackage[mathlines]{lineno}
\usepackage{hyperref}
\usepackage{bm}
\usepackage{epstopdf}
\usepackage{epsfig}
\usepackage{bbold}

\newcommand{\ket}[1]{\ensuremath{\left|#1\right\rangle}}
\newcommand{\braket}[2]{\ensuremath{\left\langle #1 | #2 \right\rangle}}

\usepackage{color}

\newcommand{\AF}[1]{\textcolor{black}{ #1}}

\begin{document}


\title{High-bit-rate quantum key distribution with entangled internal degrees of freedom of photons}

\author{Isaac~Nape}
\affiliation{School of Physics, University of the Witwatersrand, Private Bag 3, Wits 2050, South Africa}
\author{Bienvenu~Ndagano}
\affiliation{School of Physics, University of the Witwatersrand, Private Bag 3, Wits 2050, South Africa}
\author{Benjamin~Perez-Garcia}
\affiliation{School of Physics, University of the Witwatersrand, Private Bag 3, Wits 2050, South Africa}
\affiliation{Photonics and Mathematical Optics Group, Tecnol\'{o}gico de Monterrey, Monterrey 64849, Mexico}	
\author{Stirling~Scholes}
\affiliation{School of Physics, University of the Witwatersrand, Private Bag 3, Wits 2050, South Africa}
\author{Raul~I.~Hernandez-Aranda}
\affiliation{Photonics and Mathematical Optics Group, Tecnol\'{o}gico de Monterrey, Monterrey 64849, Mexico}
\author{Thomas Konrad}
\affiliation{College of Chemistry and Physics, University of KwaZulu-Natal, Private Bag X54001, Durban 4000, South Africa} 
\author{Andrew~Forbes}
\email[Corresponding author: ]{andrew.forbes@wits.ac.za}
\affiliation{School of Physics, University of the Witwatersrand, Private Bag 3, Wits 2050, South Africa}
\date{\today}

\begin{abstract}
\noindent \textbf{Quantum communication over long distances is integral to information security and has been demonstrated in free space and fibre with two-dimensional polarisation states of light. Although increased bit rates can be achieved using high-dimensional encoding with spatial modes of light, the efficient detection of high-dimensional states remains a challenge to realise the full benefit of the increased state space.  Here we exploit the entanglement between spatial modes and polarization to realise a four-dimensional quantum key distribution (QKD) protocol.  We introduce a detection scheme which employs only static elements, allowing for the detection of all basis modes in a high-dimensional space deterministically.   As a result we are able to realise the full potential of our high-dimensional state space, demonstrating efficient QKD at high secure key and sift rates, with the highest capacity-to-dimension reported to date. This work opens the possibility to increase the dimensionality of the state-space indefinitely while still maintaining deterministic detection and will be invaluable for long distance ``secure and fast" data transfer}.
\end{abstract}

\maketitle

\section{Introduction}
The use of polarization encoded qubits has become ubiquitous in quantum communication protocols with single photons \cite{Hubel2007,Ursin2007,Ma2012,Herbst2015}. Most notably, they have enabled unconditionally secure cryptography protocols through quantum key distribution (QKD) over appreciable distances \cite{Jennewein2000,Poppe2004,Peng2007}. With the increasing technological prowess in the field, faster and efficient key generation together with robustness to third party attacks have become paramount issues to address. A topical approach to overcome these hurdles is through higher-dimensional QKD: increasing the dimensionality, $d$, of a QKD protocol leads to better security and higher secure key rates, with each photon carrying up to $\log_2(d)$ bits of information \cite{bechmann2000quantum,cerf2002security}.

Employing spatial modes of light, particularly those carrying orbital angular momentum (OAM), has shown considerable improvements in data transfer of classical communication systems \cite{Wang2012, Sleiffer2012,Huang2013}. However, realizing high-dimensional quantum communication remains challenging. To date, the list of reports on high dimensional QKD with spatial modes is not exhaustive, and include protocols in up to $d=7$ \cite{Groblacher2006,mafu2013higher,mirhosseini2015high}. It is worth noting that due to experimental limitations, the secret key rate of a given QKD protocol does not scale with the dimension, i.e., given a certain set of experimental parameters there exist an optimum number of dimension that maximizes the secret key rate \cite{leach2012secure}. 

Photons with complex spatial and polarization structure, commonly known as vector modes, have been used as information carriers for polarisation encoded qubits in alignment-free QKD \cite{Souza2008,vallone2014free}, exploiting the fact that vector modes that carry OAM exhibit rotational symmetry, removing the need to align the detectors in order to reconcile the encoding and decoding bases, as would be the case in QKD with only polarization. In these vector modes, the spatial and polarization degrees of freedom (DoFs) are coupled in a non-separable manner, reminiscent of entanglement in quantum mechanics. This non-separability can be used to encode information and has been done so with classical light \cite{Milione2015e,Li2016}, for example, in mode division multiplexing \cite{Milione2015f}.   
  
Here we use the non-separability of vector OAM modes (vector vortex modes) to realize four-dimensional QKD based on the ``BB84'' protocol \cite{bennett1984quantum}. Rather than carrying information encoded in one DoF, the non-separable state can itself constitute a basis for a higher dimensional space that combines two DoFs, namely the spatial and polarisation DoFs. To fully benefit from the increased state space, we introduce a new detection scheme that, deterministically and without dimension dependent sifting loss, can detect all basis elements in our high-dimensional space.  This differs from previous schemes that have used mode filters as detectors, sifting through the space one mode at a time, thus removing all benefit of the dimensionality of the space (see for example ref.  \cite{mafu2013higher}).
Our approach combines manipulations of the dynamic and Panchanratnam-Berry phase with static optical elements and, in principle, allows detection of the basis elements with unit probability.  We demonstrate high-dimensional encoding/decoding in our entangled space, obtaining a detection fidelity as high as $97\%$, with a secret key rate of $1.63$ bits per photon and quantum error rate of $3\%$. As a means of comparison to other protocols, we calculated the capacity-to-dimension ratio and show that our scheme is more efficient than any other reported to date.

\section{Results}
\begin{figure}[h!]
	\includegraphics[width=1\linewidth]{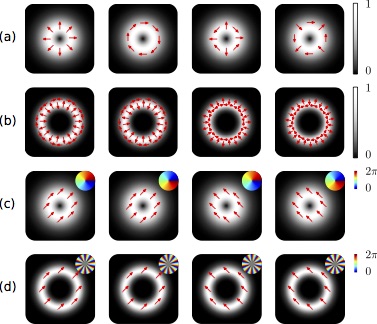}
	\caption{\textbf{Modes in a four-dimensional hyper-entangled space.} Vector vortex modes for (a) $\ell = \pm 1$ and (b) $\ell = \pm 10$, with the mutually unbiased scalar modes also for the (c) $\ell = \pm 1$ and (d) $\ell = \pm 10$ subspaces.  The insets show the azimuthally varying phase profile of the scalar modes.}		
	\label{fig: basis states}
\end{figure}

\begin{figure*}[t]
	\centering
	\includegraphics[width=\linewidth]{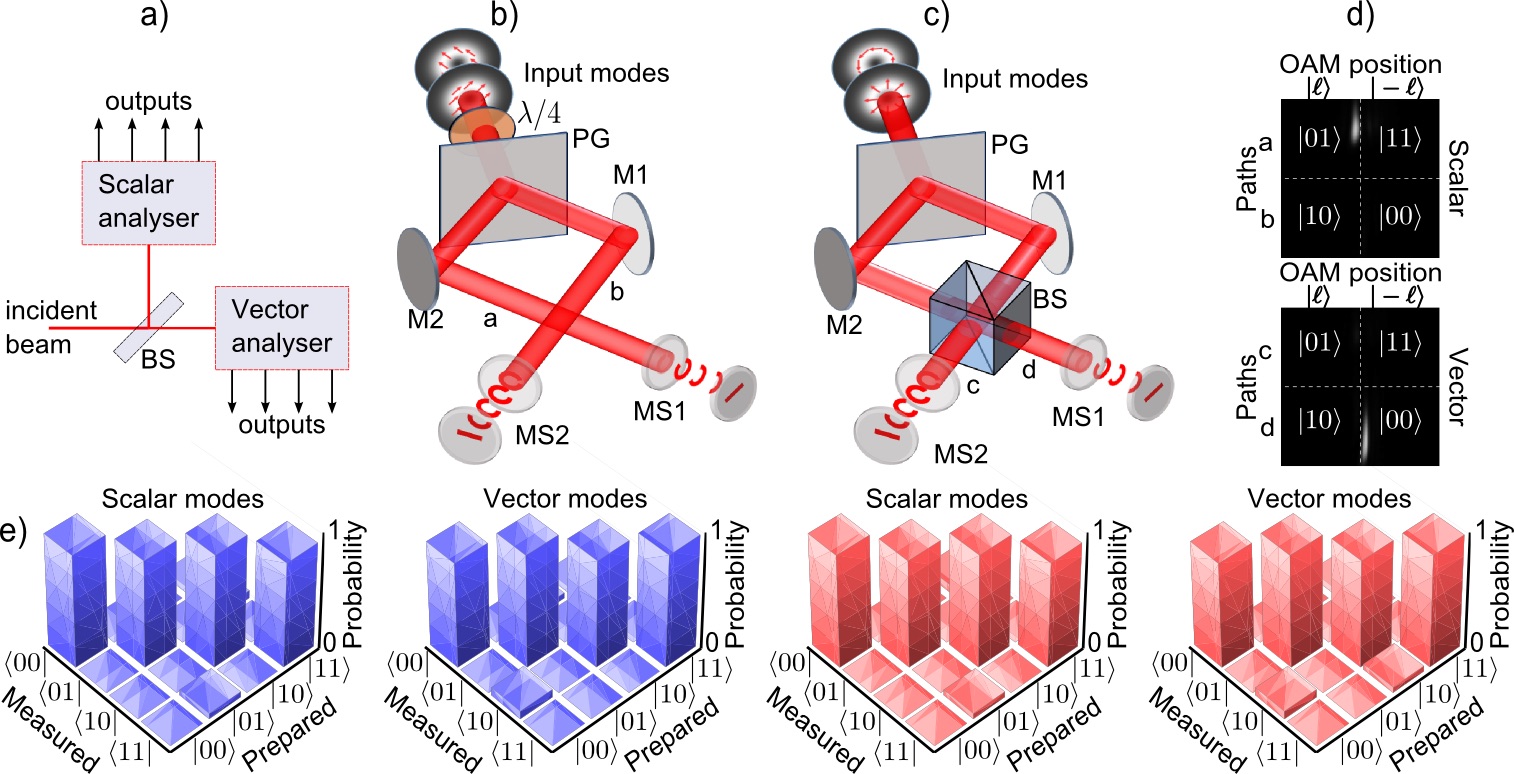}
	\caption{\textbf{Deterministic detection of the full state space.}  (a) Bob randomly selects to measure the incoming single photon from Alice in either the $\ket{\psi}_{\ell}$ (vector) or $\ket{\phi}_{\ell}$ (scalar) basis and detects the photon deterministically with eight detection ports. (b) The $\ket{\phi}_{\ell}$ scalar analyser first convert linear to circular polarisation using a quarter-wave plate, then polarisation to path with a polarisation grating (PG). Subsequently the OAM states are measured using mode sorters (MS1 and MS2) that map OAM to position. (c) The $\ket{\psi}_{\ell}$ vector analyser works in analogous fashion, with the exception that the paths after the PG are interfered on a beam-splitter (BS) before passing the resulting output from each port to an OAM detector (mode sorter). (d) The inputs states prepared by Alice are now unambiguously mapped to detectors in Bob's measurement system, allowing all states to be detected with the eight detectors.  Experimental data is shown for two of the cases.  (e) Experimental confirmation of Bob's detection scheme for both $\ket{\psi}_{\ell}$ and $\ket{\phi}_{\ell}$ states prepared by Alice, for $\ell = \pm 1$ (blue) and $\ell = \pm 10$ (red).}
	\label{fig:figure2}
\end{figure*}

\textbf{High-dimensional encoding.} The first QKD demonstrations were performed using the polarisation DoF, namely, states in the space spanned by left- circular $\ket{L}$ and right-circular polarization $\ket{R}$, i.e.,  ${\cal{H}_{\sigma}}= \mbox{span}\{\ket{L}, \ket{R}\}$, and later using the spatial mode of light as a DoF, e.g., space spanned by the OAM modes $\ket{\ell}$ and $\ket{-\ell}$, i.e. $\cal{H}_{\ell} =  \mbox{span}\{\ket{\ell}, \ket{-\ell}\}$.
Using entangled states in both DoFs allows one to access an even larger state space, i.e., ${\cal H}_{\Omega} = \cal{H}_{\sigma}\otimes\cal{H}_{\ell}$, described by the higher-order Poincar\'e sphere \cite{Milione2011a,Milione2012a}.  When many $\ell$ subspaces are combined, the dimension $d$ of the final space incorporating $N$ OAM values ($\ell  \in \Omega \subset \mathbf{N}$) is given by $d=4N$.  This opens the way to infinite dimensional encoding using such hyper-entangled states.


For example, using only the $|\ell| $ subspace of OAM ($N = 1$)  leads to a four dimensional space spanned by $\{ \ket{\ell,L},\ket{-\ell,L}, \ket{\ell,R},\ket{-\ell,R} \} $. It is precisely in this four-dimensional subspace that, here, we define our vector and scalar modes.  Alice randomly prepares photons in modes from two sets: a vector mode set, $\ket{\psi}_{\ell,\theta}$, and a mutually unbiased scalar mode set, $\ket{\phi}_{\ell,\theta}$, defined as
\begin{eqnarray}
\ket{\psi}_{\ell,\theta} &=& \frac{1}{\sqrt{2}}(\ket{R}\ket{\ell} + e^{i\theta} \ket{L}\ket{-\ell}), \label{eq: vector mode 1}\\
\ket{\phi}_{\ell,\theta} &=& \frac{1}{\sqrt{2}}\left(\ket{R} + e^{i\left(\theta-\frac{\pi}{2}\right)} \ket{L}\right)\ket{\ell}, \label{eq: scalar mode 1}
\end{eqnarray}
\noindent where each photon carries $\ell\hbar$ quanta of OAM, $\ket{R}$ and $\ket{L}$ are, respectively, the right and left circular polarization eigenstates and $\theta = [0,\pi]$ is the intra-modal phase. For a given $|\ell|$ OAM subspace, there exist four orthogonal modes in both the vector basis (Eq.~\ref{eq: vector mode 1}) and its mutually unbiased counterpart (Eq.~\ref{eq: scalar mode 1}), such that $\left|\braket{\psi}{\phi}\right|^2 = 1/d$ with $d=4$. These vector and scalar modes can be generated by manipulating the dynamic or geometric phase of light \cite{Forbes2016,Ndagano2016,Naidoo2016,Lu2016}.
\AF{Here we employ geometric phase control through a combination of $q$-plates \cite{Marrucci2006,marrucci2011spin}
and wave plates to create all vector and scalar modes in the four dimensional space (see Methods and Supplementary Information).  Our four vector modes for QKD then become:}

\begin{eqnarray}
\ket{00} &=& \frac{1}{\sqrt{2}}(\ket{R}\ket{\ell} + \ket{L}\ket{-\ell}), \\
\ket{01} &=& \frac{1}{\sqrt{2}}(\ket{R}\ket{\ell} - \ket{L}\ket{-\ell}), \\
\ket{10} &=& \frac{1}{\sqrt{2}}(\ket{L}\ket{\ell} + \ket{R}\ket{-\ell}), \\
\ket{11} &=& \frac{1}{\sqrt{2}}(\ket{L}\ket{\ell} - \ket{R}\ket{-\ell}), 
\end{eqnarray}

with corresponding MUBs 
\begin{eqnarray}
\ket{00} &=& \frac{1}{\sqrt{2}}\ket{D}\ket{-\ell}, \\
\ket{01} &=& \frac{1}{\sqrt{2}}\ket{D}\ket{\ell}, \\
\ket{10} &=& \frac{1}{\sqrt{2}}\ket{A}\ket{-\ell}, \\
\ket{11} &=& \frac{1}{\sqrt{2}}\ket{A}\ket{\ell}, 
\end{eqnarray}
where $D$ and $A$ refer to diagonal and anti-diagonal polarisation states. For the purpose of demonstration, we use vector and scalar modes in the $\ell = \pm 1$ and $\ell = \pm 10$ OAM subspaces, shown graphically in Fig.~\ref{fig: basis states}.

\textbf{High-dimensional decoding.} At the receiver's end, Bob randomly opts to measure the received photon in either the scalar or vector basis. The randomness of the choice between the two bases is implemented here with a 50:50 beam splitter (BS) as shown in Fig.~\ref{fig:figure2}(a).  Prior QKD experiments beyond two-dimensions have used filtering based techniques that negate the very benefit of the increased state space: by filtering for only one mode at a time, the effective data transfer rate is reduced by a factor $1/d$.  We introduce a new scheme to deterministically detect the modes, as detailed in Fig.~\ref{fig:figure2} (b) and (c), that has a number of practical advantages for quantum cryptography.  Consider a vector mode as defined in Eq.~\ref{eq: vector mode 1}. The sorting of the different vector modes is achieved through a combination of geometric phase control and multi-path interference.  First, a polarisation grating based on geometric phase acts as a beam splitter for left- and right-circularly polarised photons, creating two paths
\begin{equation}
\ket{\Psi}_{\ell,\theta} \rightarrow \frac{1}{\sqrt{2}} \left( \ket{\ell}_a\ket{R}_a + e^{i\theta} \ket{-\ell}_b\ket{L}_b \right),
\label{eq: vector mode2}
\end{equation}
where the subscript $a$ and $b$ refer to the polarisation-marked paths.
The photon paths $a$ and $b$ are interfered at a 50:50 BS, resulting in the following state after the BS:
\begin{equation}
\ket{\Psi'}_{\ell,\theta} = \frac{1+e^{i(\delta+\theta+\frac{\pi}{2})}}{2} \ket{\ell}_{c} + i\frac{1+e^{i(\delta+\theta-\frac{\pi}{2})}}{2} \ket{-\ell}_{d}
\label{eq: path interference}
\end{equation}
where the subscripts $c$ and $d$ refer to the output ports of the beam splitter and $\delta$ is the dynamic phase difference between the two paths. Note that the polarisation of the two paths is automatically reconciled in each of the output ports of the beam splitter due to the difference of parity in the number of reflections for each input arm. Also note that at this point it is not necessary to retain the polarisation kets in the expression of the photon state since the polarisation information is contained in the path.
In our setup we set $\delta = \pi/2$, reducing the state in Eq. \ref{eq: path interference} to 

\begin{equation}
\ket{\Psi'}_{\ell,\theta} = \frac{1-e^{i\theta}}{2} \ket{\ell}_{c} + i\frac{1+e^{i\theta}}{2} \ket{-\ell}_{d} 
\label{eq: path interference2}
\end{equation}
The measurement system is completed by passing each of the outputs in $c$ and $d$ through a mode sorter and collecting the photons using 4 multimode fibres coupled to avalanches photodiodes.  The mode sorters are refractive (lossless) aspheres that map OAM to position \cite{Berkhout2010a, Fickler2014a, Lavery2013, Dudley2013} (see Supplementary Information for a layout of the detection system).   While it is trivial to measure such hyper-entangled (non-separable) vector states at the classical level \cite{Milione2015e,Milione2015f,Ndagano2015},
with our approach each such state is detected with unit probability at the single photon level.  For example, consider the modes $\ket{00}$ and $\ket{01}$, where $\theta = 0$ and $\theta = \pi$, respectively.  The mapping is such that 

\begin{eqnarray}
\ket{00} &\rightarrow& \ket{\Psi'}_{\ell,0} = i \ket{-\ell}_{d}, \\ 
\ket{01} &\rightarrow& \ket{\Psi'}_{\ell,\pi} = - \ket{\ell}_{c}. 
\label{eq:mapping}
\end{eqnarray}

\noindent The combination of path ($c$ or $d$) and lateral location ($+\ell$ or $-\ell$) uniquely determines the original vector mode as shown in Fig.~\ref{fig:figure2}(d)

The scalar mode detector works on an analogous principle but without the need of the BS to resolve the intermodal phases (see Supplementary Information). The polarisation states are resolved by first performing a unitary transformation that maps linear to circular basis, and passing the scalar mode through the polarisation grating. The OAM states are subsequently sorted using the mode sorters.

A graphical illustration of the experimental performance of both the scalar and vector analysers is shown in Fig.~\ref{fig:figure2}(e), where modes from the $\ell = \pm 1$ and $\ell = \pm 10$ subsets were measured with high fidelity (close to unity).
%
%

\begin{figure}[t]
	\includegraphics[width=\linewidth]{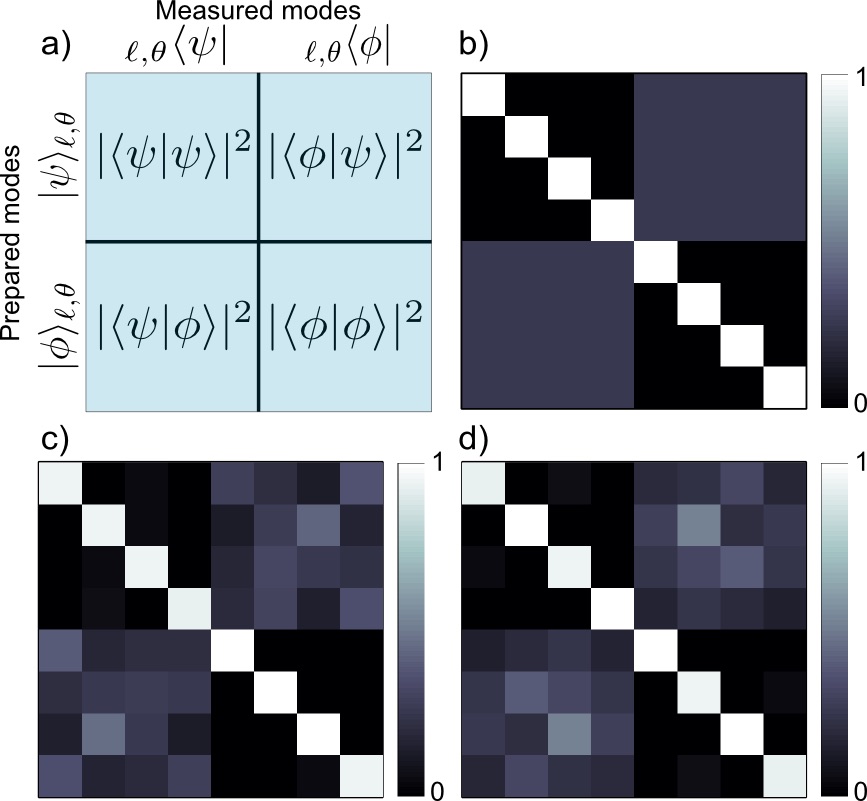}
	\caption{\textbf{Crosstalk analysis in four dimensions.}  (a) Schematic of the inner product measurements performed between the vector states $\ket{\psi}_{\ell,\theta}$ and their mutually unbiased counterparts $\ket{\phi}_{\ell,\theta}$. (b) Theoretical scattering probabilities among the vector and scalar modes following the measurement process of (a). The experimental results are shown in (c) and (d) for modes in the $\ell = \pm 1$ and $\ell = \pm 10$ subspaces, respectively.}
	\label{fig: figure3}
\end{figure}
\textbf{High dimensional cryptography.} We performed a four-dimensional prepare-and-measure BB84 scheme \cite{bennett1984quantum} using mutually unbiased vector and scalar modes. Light from our source was attenuated to the single photon level with an average photon number of $\mu = 0.008$.  Alice prepared an initial state in either the $\ket{\psi}_{\ell}$ (vector) or $\ket{\phi}_{\ell}$ (scalar) basis and transmitted it to Bob, who made his measurements as detailed in the previous section.  Through optical projection onto both the vector and scalar bases as laid out in Fig.~\ref{fig: figure3}(a), we determined the crosstalk matrices shown in Fig.~\ref{fig: figure3}(c) and (d), relating the input and measured modes within, respectively, the subspaces $\ell = \pm 1$ and $\ell = \pm 10$. The average fidelity of detection, measured for modes prepared and detected in identical bases, is $0.965\pm0.004$ while the overlap between modes from MUBs is $|\braket{\phi}{\psi}|^2 = 0.255 \pm 0.004$, in good agreement with theory (0.25).

From the measured crosstalk matrices in Fig.~\ref{fig: figure3}(c) and (d), we performed a security analysis on our QKD scheme in dimensions $d=4$ for the two OAM subspaces ($\pm 1$ and $\pm 10$). The results of the analysis are summarised in Table \ref{tab:table2}. From the measured detection fidelity $F$, we computed the mutual information between Alice and Bob in $d$-dimensions as follows \cite{cerf2002security}
\begin{equation}
I_{AB}=\log_{2}(d) + F\log_{2}(F)+ (1-F)\log_{2}\left(\frac{1-F}{d}\right).
\end{equation}
The measured $I_{AB}$ for $d=4$ is nearly double ($1.7 \times$) that of the maximum achievable with only qubit states ($1$).
Assuming a third party, Eve, uses an ideal quantum cloning machine to extract information, the associated cloning fidelity, $F_E$, in $d$-dimensions is given by \cite{cerf2002security}
\begin{equation}
F_{E}=\frac{F}{d}+\frac{(d-1)(1-F)}{d} + \frac{2\sqrt{(d-1)F(1-F)}}{d}.
\end{equation}
With increasing dimensions, the four dimensional protocol reduces the efficiency of Eve's cloning machine to as low as $0.41$ well below the maximum limit in a the two-dimensional protocol (0.5)
Thus, increasing the dimensionality of QKD protocols does indeed have, in addition to higher mutual information capacity, higher robustness to cloning based attacks.

The mutual information shared between Alice and Bob, conditioned on Bob's error -- that is, Bob making a wrong measurement is as a result of Eve extracting the correct information -- is computed in $d$-dimension as follows \cite{cerf2002security}
\begin{equation}
\begin{split}
I_{AE}= & \log_{2}(d) + (F+F_E-1)\log_{2}\left(\frac{F+F_E-1}{F}\right)\\ & +(1-F_E)\log_{2}\left(\frac{1-F_E}{(d-1)F}\right).
\end{split}
\end{equation}
\begin{table}
	\centering
	\caption{Summary of the security analysis on the high dimensional protocol showing the experimental and theoretical values of the detection fidelity ($F$), mutual information $I_{AB}$ between Alice and Bob, Eve's cloning fidelity $(F_{E})$ and mutual information with Alice $I_{AE}$, as well as the quantum error rate $Q$ and secret key rate $R$.}
	\label{tab:table2}
	\begin{tabular}{|c|c|c|c|}
		\hline 
		&  $d=4$ ($|\ell|$ = 1)  & $d=4$ ($|\ell|$ = 10) &  \\ 
		\hline 
		Measures 	& experiment & experiment & ideal \\ 
		\hline 
		$F$			& $0.96$ & $0.97$ &  $1.00$\\ 
		\hline 
		$I_{AB}	$	& $1.69$ & $1.76$ & $2.00$ \\ 
		\hline 
		$F_{E}	$	& $0.44$ & $0.41$ & $0.25$ \\ 
		\hline 
		$I_{AE}	$	& $0.17$ & $0.13$	 & $0.00$ \\ 
		\hline 
		$Q$			& $0.04$ & $0.03$  & $0.00$ \\ 
		\hline 
		$R$			& $1.52$ & $1.63$ & $2.00$ \\ 
		\hline 
	\end{tabular} 
\end{table}

\begin{figure*}[t]
	\centering
	\includegraphics[width=\linewidth]{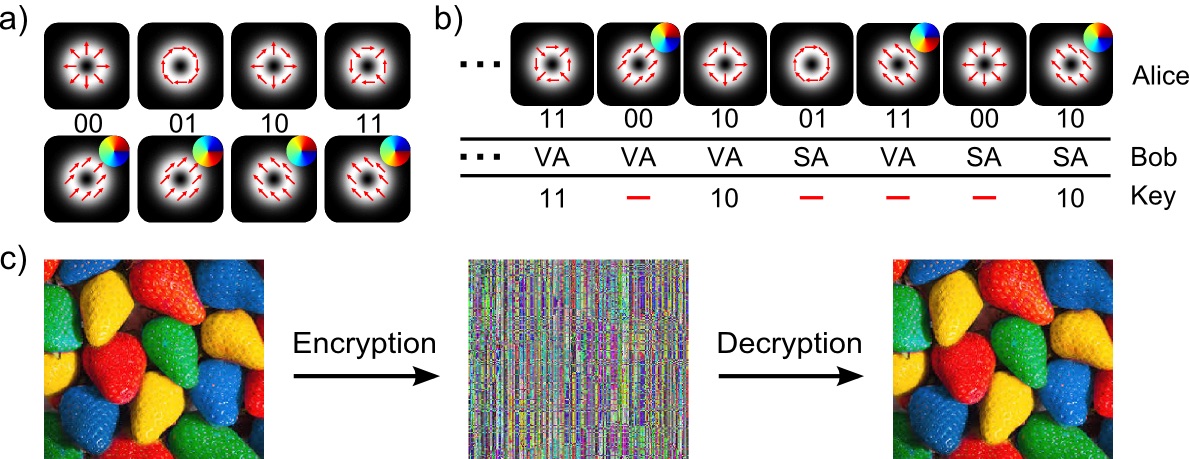}
	\caption{\textbf{High dimensional BB84.} Alice and Bob agree on bit values for the vector and scalar modes. (b) Alice sends a random sequence of vector and scalar modes, which Bob randomly measures using either a vector analyser (VA) or a scalar analyser (SA). Alice and Bob, upon communication of the encoding and decoding bases through a classical channel, discard bit values for modes prepared and measured in complementary bases. (c) Shows a simple encryption/decryption of an image using a 98 bit long key, sifted from a total of 200 transmitted bits.}
	\label{fig:figure4}
\end{figure*}

The consequent measured quantum error rate of $Q = 1-F = 0.04$ is well below the $0.11$ and $0.18$ bounds for unconditional security against coherent attacks in two and four dimensions \cite{cerf2002security}, respectively. The lower bound on the secret key rate, $R=\max\left(I_{AB} - I_{AE},I_{AB} - I_{BE}\right)$ \cite{gisin2002quantum}, yields a value as high as $1.63$ bits per photon, well above the Shannon limit of one bit per photon achievable with qubit states. While the security of the protocol can be increase with privacy amplification, the measured four-dimensional secret key rate demonstrates the potential of such hyper-entangled modes for high bandwidth quantum communication.

Finally, we performed a four dimensional prepare-and-measure BB84 scheme using mutually unbiased vector and scalar modes. For each mode, Alice and Bob assign the bit values $00, 01, 10 $ and $11$, as shown in Fig.~\ref{fig:figure4}(a).
During the transmission, Alice randomly prepares her photon in a vector (scalar) mode state while Bob randomly measures the photon with either the vector or scalar analyser detailed in Fig. \ref{fig:figure2}. At the end of the transmission, Alice and Bob reconcile the prepare and measure bases and discard measurements in complementary bases, as described in Fig.~\ref{fig:figure4}(b). We performed this transmission using a sequence of 100 modes and retained a sifted key of 49 spatial modes (98 bits), which was used to encrypt and decrypt a picture as shown in Fig.~\ref{fig:figure4}(c).

\section{Discussion and conclusion}
The prepare-and-measure quantum cryptography scheme we report here realised the potential of hyper-entanglement between spatial modes and polarisation as means to achieving higher bandwidth optical communication at the single photon level as well as classically. Our secret key rate of 1.63 bits per photon represents a significant increase in data transfer rates as compared to  QKD with conventional polarisation eigenstates, limited to one bit per photon under ideal conditions. The secret key rate we obtained exceeds previously reported \cite{mafu2013higher} $d = 4$ laboratory results by more than $43\%$. In order to compare the efficiency of higher-dimensional protocols we define the information per photon per dimension as a figure of merit.  Using this, we find that we achieve a value of $0.41$, compared to reported values of  $0.17$ ($d = 5$) \cite{mafu2013higher} and $0.24$ ($d = 7$) \cite{mirhosseini2015high}, highlighting the efficiency of our scheme.

An important aspect of our scheme is the deterministic measurement of all higher dimensional states, allowing, in principle, unit detection probability by Bob for any prepared mode by Alice. This makes it possible to increase the dimensionality of quantum cryptography protocols without compromising on the sifting rate, the fraction of the transmitted bits that constitute the key, unlike with other methods where the data transfer rate is decreased due to filtering for one mode at a time, thereby decreasing the detection probability for a given mode by a factor $1/d$.  As a consequence our sift rate is two times greater than would be possible with conventional probabilistic (filtering-based) detection schemes (See Supplementary Information for an experimental comparison).  We point out that our scheme would likewise increase the signal-to-noise of classical mode division multiplexing communication systems: rather than distribute the signal across $d$ modes, each with $1/d$ of the signal, we can achieve full signal on each mode with a factor $d$ greater signal-to-noise ratio \cite{Ruffato2016}. 

In conclusion, we have demonstrated a four-dimensional QKD protocol using a deterministic detection scheme that realises the full benefit of of the dimensionality of the state space.  Using modes with entangled spatial and polarisation DoFs, we demonstrated the efficiency of the approach using the BB84 scheme.  The system performance confirms that the QKD protocol is capable of realising high-bits per photon at high sift rates and high data transfer rates, substantially improving on previously reported results.  It is anticipated that, due to the identical scattering of vector and scalar OAM modes in turbulence \cite{Cox2016}, no benefit will be derived to Eve (mutual information between Bob/Alice and Eve) from this mode set.  When combined with real-time error correction \cite{Ndagano2016}
and the possibility to increase the dimensionality of the state-space indefinitely while still maintaining unit probability detection, we foresee that this approach will be invaluable for long distance ``secure and fast" data transfer.

\section{Methods}

\textbf{Generating vector and scalar modes using a $q$-plate.} We used $q$-plates to couple the polarisation and orbital angular momentum degrees of freedom through geometric phase control. With locally varying birefringence across a wave plate, the geometric phase imparted by a $q$-plate was engineered to produce the following transformation
\begin{eqnarray}
\left| \ell, L \right\rangle \xrightarrow{q\text{-plate}} \left| \ell + 2q, R \right\rangle, \label{eq:Qplate1}\\
\left| \ell, R \right\rangle \xrightarrow{q\text{-plate}} \left| \ell - 2q, L \right\rangle,
\label{eq:Qplate2}
\end{eqnarray} 
where $q = 2\ell$ is the topological charge of the $q$-plate. The vector modes investigated here were generated by transforming an input linearly polarised Gaussian mode with quarter- or half- wave plates and $q=1/2$ and $q=5$ plates, producing either separable (scalar) non-separable (vector) superpositions of qubit states in Eqs. \ref{eq:Qplate1} and \ref{eq:Qplate2}.  The generated states and the elements setting are given in the table below:

\begin{table}[h!]
	\caption{Generation of MUBs of vector and scalar modes from an input, horizontally polarised Gaussian beam}
	\centering
	\begin{tabular}{|c|c|c|c|c|c|}
		\hline 
		Mode & $\lambda/4 (\alpha_1) $ & $\lambda/2 (\theta_1)$ & $q$-plate & $\lambda/4 (\alpha_2)$ & $\lambda/2 (\theta_2)$ \\
		\hline
		\hline 
		$\ket{\psi}_{\ell,0}$ & -- & 0 & $|q|$ & -- & -- \\ 
		\hline 
		$\ket{\psi}_{\ell,\pi}$ & -- & $\pi/4$ & $|q|$ & -- & -- \\ 
		\hline 
		$\ket{\psi}_{-\ell,0}$ & -- & -- & $|q|$ & -- & 0 \\ 
		\hline 
		$\ket{\psi}_{-\ell,\pi}$ & -- & -- & $|q|$ & -- & $\pi/4$ \\ 
		\hline
		\hline
		$\ket{\phi}_{\ell,0}$ & $-\pi/4$ & -- & $|q|$ & $-\pi/4$ & $\pi/4$  \\ 
		\hline 
		$\ket{\phi}_{\ell,\pi}$ & $-\pi/4$ & -- & $|q|$ & $-\pi/4$ & $-\pi/4$ \\ 
		\hline 
		$\ket{\phi}_{-\ell,0}$ & $\pi/4$ & -- & $|q|$ & $\pi/4$ & $\pi/4$ \\ 
		\hline 
		$\ket{\phi}_{-\ell,\pi}$ & $\pi/4$ & -- & $|q|$ & $\pi/4$ & $-\pi/4$ \\ 
		\hline 
	\end{tabular}
\end{table}

\section*{Materials and correspondence}
Correspondence and requests for materials should be addressed to A.F.


\section*{Acknowledgments}
We express our gratitude to Lorenzo Marrucci for providing us with $q$-plates and Miles Padgett and Martin Lavery for the mode sorters. B.N. acknowledges financial support from the National Research Foundation of South Africa and I. N. from the Department of Science and Technology (South Africa). B.P.G. and R.I.H. acknowledge support from CONACyT.



\section*{Authors' contributions}
Experiments were performed by B.N., I.N., S.S. and B.P.G.  All authors contributed to the data analysis and interpretation of the results. A.F., I.N. and B.N. wrote the manuscript with inputs from all the authors. A.F. supervised the project.



\section*{Competing financial interests}
The authors declare no financial competing interests.

\newpage

\section{Supplementary information }

\textbf{Sorting of scalar OAM mode.} We use a compact phase element to perform a geometric transformation on OAM modes such that azimuthal phase is mapped to a transverse phase variation,  i.e., a tilted wavefront. The first optical element of our OAM mode sorter performs a conformal mapping in the standard Cartesian coordinates, from a position in the input plane $(x,y)$ to one in the output plane $(u,v)$, such that
\begin{figure}[h!]
	\includegraphics[width=\linewidth]{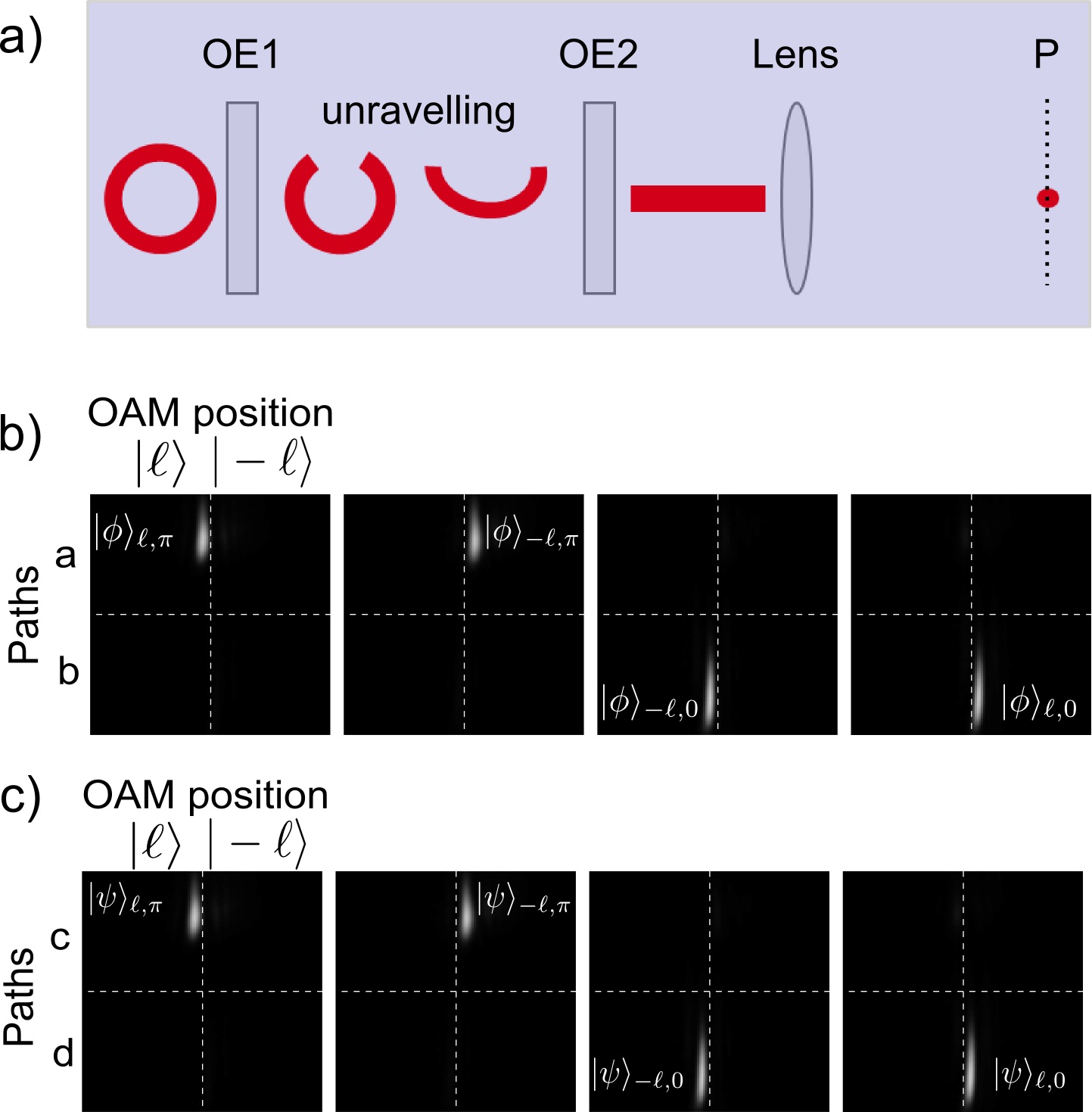}
	\caption{ \textbf{Sorting the modes}. (a)We use a mode sorter that consist of two refractive optical lenses, OE1 and OE2 which transform azimuthal phase into a linear phase and mapped onto positions . The input mode unravels after OE1 and a linear phase is retained by OE2. The phase is then mapped onto a position unique to the azimuthal charge by a Fourier lens. We use these lenses to map the modes set $(\ket{\phi}, \ket{\psi})$ onto positions based on their orbital angular momentum system illustrated in Fig.~\ref{fig:figure2} (c) and Fig.~\ref{fig:figure2} (d). }
	\label{fig:figure5}
\end{figure}
\begin{eqnarray}
u &=& \frac{d}{2\pi}\arctan\left(\frac{y}{x}\right) \\
v &=&  -\frac{d}{2\pi}\ln\left(\frac{\sqrt{x^2+y^2}}{b}\right)
\end{eqnarray}
where $d$ is the aperture size of the free form optics and $b$ is a scaling factor that controls the translation of the transformed beam in the $u$ direction of the new coordinate system.  The result is that after passing through a second phase-correcting optic and then a Fourier transforming lens (of focal length $f$), the input OAM ($\ell$) is mapped to output positions ($X_{\ell}$ following
\begin{equation}
X_{\ell} = \frac{\lambda f \ell}{d}.
\end{equation}\\
\textbf{Crosstalk analysis.} The crosstalk analysis of the vector ($\ket{\psi}$) and scalar ($\ket{\phi}$) modes is represented by a matrix of detection probabilities for each of the modes sent by Alice (rows) and measured by Bob (columns). The entries are partitioned into four quadrants: the diagonal quadrants correspond to the outcomes of measurements in matching bases while the off-diagonal show the outcomes of measurements in complementary bases.\\  
\begin{figure}[h!]
	\includegraphics[width=\linewidth]{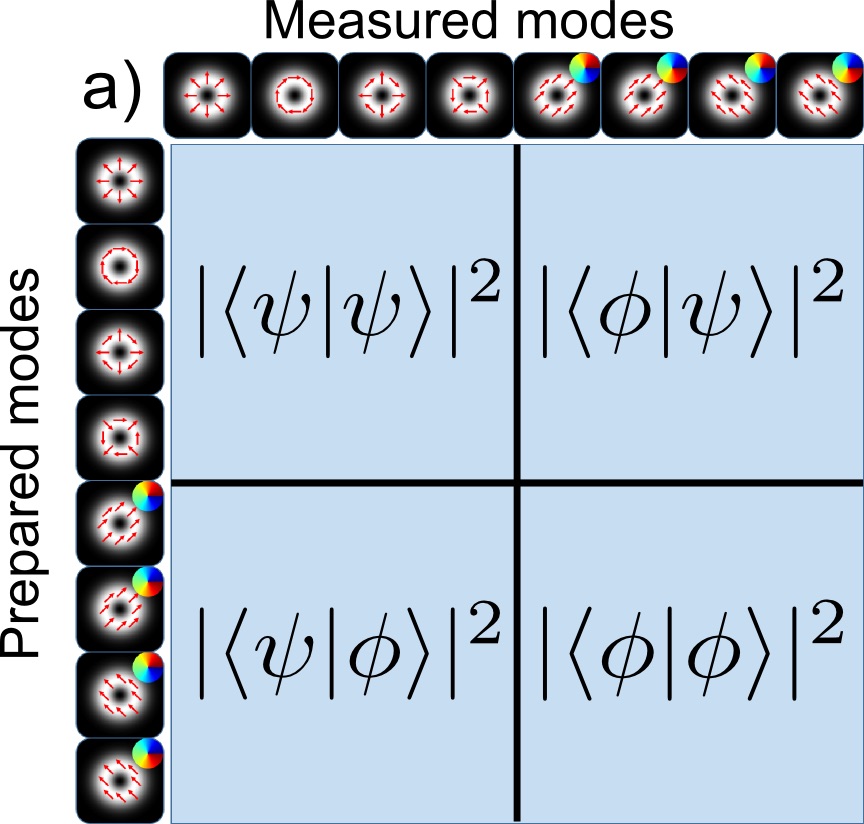}
	\caption{\textbf{Crosstalk analysis matrix for the four dimensional sets of vector and scalar modes.}  The matrix can be broken into four quadrants representing probabilities for preparation and measurements in the two basis. 
		\label{fig:figure6}}
\end{figure}	

\begin{figure}
	\includegraphics[width=\linewidth]{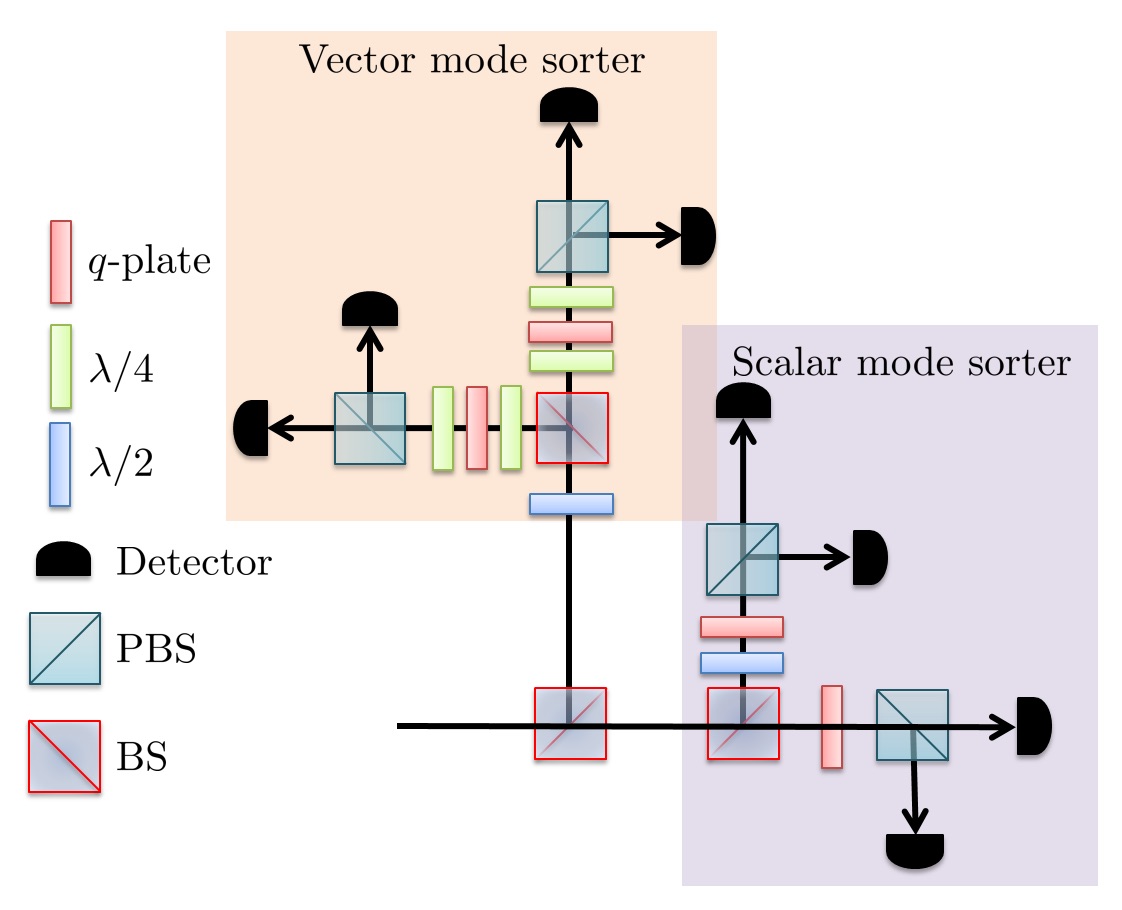}
	\caption{ \textbf{Filter based system for detecting the vector and scalar modes.} The combination of wave-plates ($\frac{\lambda}{4}$ , $\frac{\lambda}{2}$), q-plates and 50/50 (BS) and polarisation (PBS) beam splitters can serve as a detection system. The wave plates can be rotated at angles shown in Table.1  }\label{fig:figure7}
\end{figure}

\textbf{Filter based detection system.} The filter based detection system depends on the use of beam splitters with a combination with $q$-plates, wave plates and polarisers. While it is common practice for the measurement process to be identical to the generation for reversible processes -- as is the case in linear optics -- this approach would fail in measuring high dimensional vector mode spaces. This is because vector modes within one subset required oppositely charged $q$-plates. The best approach to probe the high dimensional space would require the use of beam splitters as shown in Fig.~\ref{fig:figure7}, however, at the cost of reducing the detection probability by a factor of $1/2$, thus halving the sift rate and secure key rate; for a key that is $N$-bit long, one would require sending, on average, $4N$ bits.  We have tested this by building the system depicted in Fig.~\ref{fig:figure7} and performing the same prepare and measure QKD protocol as detailed in the main text.  For a 200 bit transmission we were only able to produce a key with $25\%$ of the transmitted bits, as compared to $50\%$ using the scheme described in Fig.~\ref{fig:figure2} of the main text.  This highlights one of the advantages of a deterministic detection system versus the probabilistic (filter-based) system.

\end{document}